\documentclass[aps,prl,twocolumn,showpacs,superscriptaddress]{revtex4}
\usepackage{graphicx}
\usepackage{amsmath}
\usepackage{amssymb}
\usepackage{colordvi}
\usepackage{mathrsfs}
\usepackage{bm}
\usepackage{verbatim}
\usepackage{dcolumn}
\usepackage{bm}
\usepackage{epsfig}
\usepackage{subfigure}

\begin{document}

\title{Realization of Quantum Anomalous Hall Effect in Graphene from \textit{n}-\textit{p} Codoping Induced Stable Atomic-Adsorption}
\author{Xinzhou Deng}
\affiliation{ICQD, Hefei National Laboratory for Physical Sciences at Microscale, and Synergetic Innovation Center of Quantum Information and Quantum Physics, University of Science and Technology of China, Hefei, Anhui 230026, China.}
\affiliation{CAS Key Laboratory of Strongly-Coupled Quantum Matter Physics, and Department of Physics, University of Science and Technology of China, Hefei, Anhui 230026, China.}
\author{Shifei Qi}
\affiliation{School of Chemistry and Materials Science, Shanxi Normal University, Linfen, Shanxi 041004, China}
\affiliation{ICQD, Hefei National Laboratory for Physical Sciences at Microscale, and Synergetic Innovation Center of Quantum Information and Quantum Physics, University of Science and Technology of China, Hefei, Anhui 230026, China.}
\author{Yulei Han}
\affiliation{ICQD, Hefei National Laboratory for Physical Sciences at Microscale, and Synergetic Innovation Center of Quantum Information and Quantum Physics, University of Science and Technology of China, Hefei, Anhui 230026, China.}
\affiliation{CAS Key Laboratory of Strongly-Coupled Quantum Matter Physics, and Department of Physics, University of Science and Technology of China, Hefei, Anhui 230026, China.}
\author{Kunhua Zhang}
\affiliation{ICQD, Hefei National Laboratory for Physical Sciences at Microscale, and Synergetic Innovation Center of Quantum Information and Quantum Physics, University of Science and Technology of China, Hefei, Anhui 230026, China.}
\affiliation{CAS Key Laboratory of Strongly-Coupled Quantum Matter Physics, and Department of Physics, University of Science and Technology of China, Hefei, Anhui 230026, China.}
\author{Xiaohong Xu}
\email[Correspondence to:~~]{xuxh@dns.sxnu.edu.cn}
\affiliation{School of Chemistry and Materials Science, Shanxi Normal University, Linfen, Shanxi 041004, China}
\author{Zhenhua Qiao}
\email[Correspondence to:~~]{qiao@ustc.edu.cn}
\affiliation{ICQD, Hefei National Laboratory for Physical Sciences at Microscale, and Synergetic Innovation Center of Quantum Information and Quantum Physics, University of Science and Technology of China, Hefei, Anhui 230026, China.}
\affiliation{CAS Key Laboratory of Strongly-Coupled Quantum Matter Physics, and Department of Physics, University of Science and Technology of China, Hefei, Anhui 230026, China.}

\begin{abstract}
Using first-principles calculation methods, we study the possibility of realizing quantum anomalous Hall effect in graphene from stable 3\textit{d}-atomic adsorption via charge-compensated \textit{n}-\textit{p} codoping scheme. As concrete examples, we show that long-range ferromagnetism can be established by codoping 3\textit{d} transition metal and boron atoms, but only the Ni codopants can open up a global bulk gap to harbour the quantum anomalous Hall effect. Our estimated ferromagnetic Curie transition temperature can reach over 10 Kelvin for various codoping concentrations.
\end{abstract}

\pacs{73.20.At;
          73.43.-f,   
          75.70.Tj.
          }
\maketitle
\textit{Introduction---.} Quantum anomalous Hall effect (QAHE) is a quantized response of transverse charge current to an electric field in the absence of magnetic field~\cite{Haldane,QAHE-Rev1,QAHE-Rev2,QAHE-Rev3}. It originates from the joint effect of spin-orbit coupling and local magnetization. The presence of linear Dirac dispersion has made both graphene~\cite{Graphene} and topological insulators (TIs)~\cite{TopologicalInsulator-1,TopologicalInsulator-2} ideal platforms to explore QAHE. Comparing with graphene, TIs show great superiority because of their intrinsic spin-orbit couplings, which indicate that the only condition to realize QAHE in TIs is to break time-reversal invariance. By doping magnetic atoms into TI thin films (e.g., doping Cr/V atoms in (Bi,Sb)$_2$Te$_3$), QAHE has been theoretically proposed~\cite{TI-QAHE-Theory1,TI-QAHE-Theory2} and later experimentally observed at extremely low temperatures (e.g. $<$ 100 mK)~\cite{TI-QAHE-Exp1,TI-QAHE-Exp2,TI-QAHE-Exp3,TI-QAHE-Exp4,TI-QAHE-Exp5,TI-QAHE-Exp6,TI-QAHE-Exp7}. A charge-compensated $n$-$p$ codoping scheme was adopted to increase the QAHE observation temperature in TI thin films,  e.g., a temperature over 50 Kelvin can be achieved by codoping vanadium-iodine into Sb$_2$Te$_3$ thin films~\cite{HiT-QAHE}.

Alternatively, although graphene is nonmagnetic and exhibits extremely weak spin-orbit couplings~\cite{Graphene,WeakSOC1,WeakSOC2,WeakSOC3}, it still attracts broad attention to produce QAHE because of its superior electronic properties and broad perspective for future industry-scale applications. In Refs.~\cite{Gra-QAHE1,Gra-QAHE15,Gra-QAHE2,Gra-QAHE3}, it theoretically shows that Rashba spin-orbit coupling together with ferromagnetism lead to the formation of QAHE in graphene by periodically doping magnetic atoms. However, precise control of doping position is out of current experimental technique. This makes this QAHE proposal unrealistic because of the potentially induced inter-valley scattering during non-periodic doping scheme. Strikingly, following-up study~\cite{Gra-QAHE4} shows that random distribution of magnetic dopants in graphene can greatly eliminate inter-valley scattering. Subsequent studies from both theory and experiment showed that magnetic atoms in graphene tend to nucleate into clusters on graphene owing to the small binding energies~\cite{DopingCluster1,DopingCluster2}.
After some efforts, a rewarding approach was theoretically proposed to realize QAHE by placing graphene on magnetic insulator thin films~\cite{MagneticInsulator-theory1,MagneticInsulator-theory2}. Soon, it was experimentally reported that sizable AHE can be observed in graphene by considering YIG magnetic thin films~\cite{MagneticInsulator-exp1}. Such kind of scheme can lead to considerable ferromagnetism in graphene, but extremely weak Rashba spin-orbit coupling~\cite{MagneticInsulator-exp1,MagneticInsulator-exp2}. The fundamental reason of the weak Rashba spin-orbit coupling arises from the large separation ($\sim$ 2.5-3 {\AA}, see Refs.~\cite{MagneticInsulator-theory1,MagneticInsulator-theory2}) between graphene and magnetic insulating substrate due to weak van der Waals interaction, which shows a clear contrast with the chemical interaction in the atomic adsorption situation ($\sim$ 1.6 {\AA}, see Refs.~\cite{Gra-QAHE1,Gra-QAHE15,Gra-QAHE3}). Thus, the extremely weak Rashba spin-orbit coupling is the only ingredient to hinder the realization of QAHE.

To our knowledge, the most rewarding approach to increase Rashba spin-orbit coupling in graphene is to design a stable atomic-adsorption scheme. \Blue{Inspired by the advantages of manipulating band gaps from $n-p$ codoping method~\cite{yamamoto1999solution,wang2003cluster,gai2009design,zhu2009band,xu2006carrier, zhu2008dopant,zhang2015long,Qi-Carbon} and the successful fabrication of boron (B)-substituted graphene~\cite{BdopeMethod}}, we present a systematic investigation on the possible realization of QAHE in graphene by utilizing $n$-$p$ codoping technique. As examples, we take 3$d$ transition metal atoms as $n$-type dopants that are adsorbed on top of graphene and B-atoms as $p$-type dopants that substitute carbon atoms. We show that although the magnetic states for most 3$d$ adatoms (except Cr and Mn) show RKKY-type spatial fluctuation, only V or Ni atoms can exhibit long-range ferromagnetic orders in B-substituted graphene. When the spin-orbit coupling is further invoked, we find that only Ni-B codoping can open up a global band gap that harbours the QAHE. This is completely distinct from the situation of single-element Ni adsorbed graphene~\cite{Gra-QAHE15,Cohen} that induces vanishing ferromagnetism arising from the electron redistribution, i.e., $3d^84s^2 \rightarrow 3d^{10}4s^0$. We further show the dependence of the QAHE gap on the codoping concentration.

\textit{Computational Methods---.} The first-principles calculations were performed within the framework of density functional theory using the projected-augmented-wave method~\cite{PAW} as implemented in the Vienna Ab-initio Simulation Package~\cite{VASP1,VASP2,VASP3}. The generalized gradient approximation~\cite{GGA} of Perdew-Burke-Ernzerhof~\cite{PBE} is adopted to treat the exchange correlation interactions. To avoid the intervalley scattering in the codoping scheme, a 8$\times$8 graphene supercell is adopted. The vacuum space of 20 \AA\ between two graphene layers is set to avoid spurious interactions between periodic images. The kinetic energy cutoff and the energy convergence threshold are set to be 400 eV and $10^{-4}$ eV, respectively. The atomic structures were fully relaxed till the Hellmann-Feynman force on each ion is less than 0.02 eV/\AA. The Gaussian smearing method with a smearing width of 0.1 eV and a k-mesh point grid of 3$\times 3 \times1$ are adopted during the structural relaxation. A k-mesh point of 5$\times$5$\times$1 is used for the total energy estimation, while a $11 \times 11 \times1$ k-mesh point mesh is used in calculating density of states.

\begin{figure}
\includegraphics
[width=8cm,angle=0]{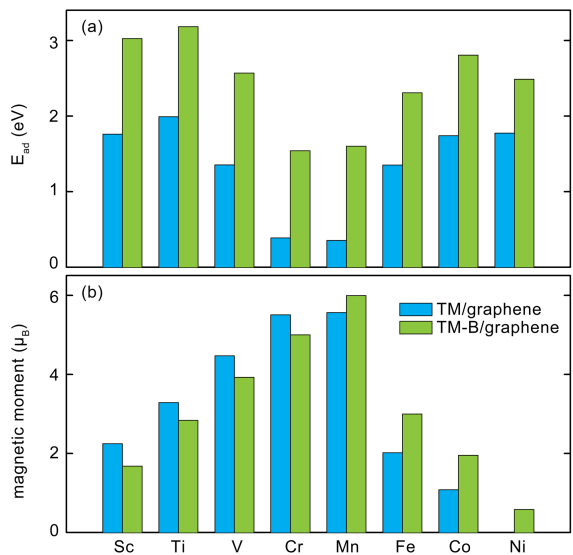}
\caption{ (color online). (a) Adsorption energy and (b) Magnetic moment of 3d TM adatoms on pristine graphene (blue) and B-codoped graphene (green).
}
\label{figure1}
\end{figure}

\textit{Adsorption Analysis---.} We begin from revisiting the single-element adsorbed graphene. Previous studies showed that most 3d metal adatoms favour hollow position~\cite{hollow1,hollow2,Gra-QAHE15}. Thus, the corresponding adsorption energy $E_{\rm{ad}}$ can be defined as:
\begin{eqnarray}
 E_{\rm{ad}} = E_{\rm{TM}} + E_{\rm{Gra}} - E_{\rm{Tot}}, \nonumber
\end{eqnarray}
where $E_{\rm{TM}}$, $E_{\rm{Gra}}$, and $E_{\rm{Tot}}$ are respectively energies of the isolated transition metal atoms, pristine graphene, and the atom-adsorbed graphene. The blue bars in Fig.~\ref{figure1}(a) display the adsorption energy of only metal atom-adsorbed graphene, which agrees well with the previous studies in the range of 0.4$-$1.9 eV~\cite{EadFeCoNi,Gra-QAHE15}. The relatively weak adsorption energy cannot prevent the fast migration of adatom-clustering when other adatoms are further adsorbed on graphene. To overcome the difficulty of forming a dilute adatom-distribution, \textit{n}-\textit{p} codoping scheme was proposed in a previous work where codoping with B atoms is able to significantly suppress those undesirable effects and the metal adatoms are closely located near the substituted B positions due to strong electrostatic attraction~\cite{Qi-Carbon}. When B codopants are considered, as displayed by green bars in Fig.~\ref{figure1}(a), the adsorption energy exhibits a $\sim$1 eV enhancement compared with single-element doping, which can effectively prevent the adatom migration and clustering on B-substituted graphene. Based on the Arrhenius equation, the room temperature corresponds to an adsorption energy of $E_{\rm{ad}}$ $\sim$0.75 eV. Thus, the estimated stable temperature of transition metal adatoms on B-doped graphene is within the range of 617 -1273 K, which is larger than that (142 K -797 K) of only adsorbing transition metal atoms on graphene. This indicates that the \textit{n}-\textit{p} codoping adsorption on graphene is more stable above the room temperature in this study. This codoping scheme provides a basic material structure with stable dilute-distribution of adatoms in graphene.

\begin{figure}
\includegraphics
[width=8.5cm,angle=0]{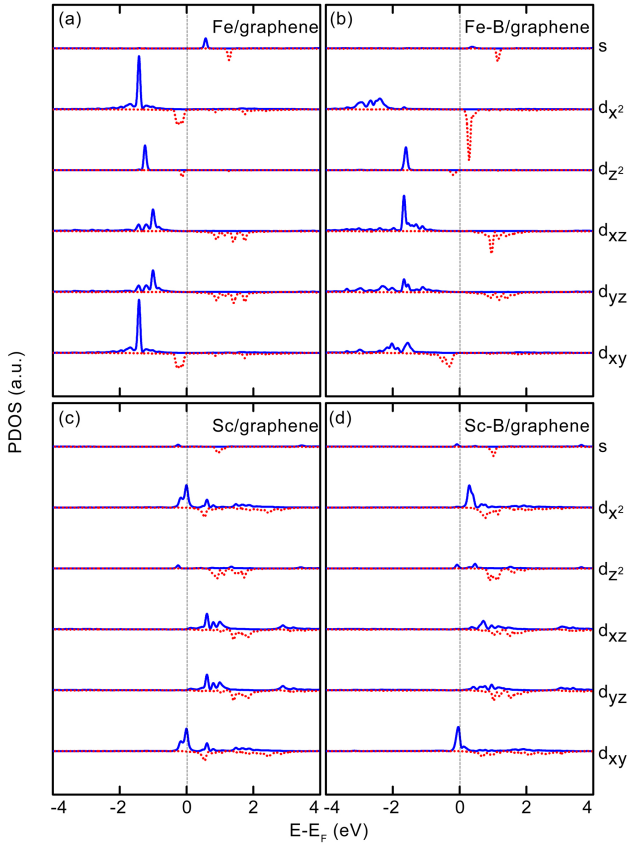}
\caption{ (color online). Orbital resolved DOS of Fe adatom on (a) pristine graphene and (b) B-codoped graphene. Orbital resolved DOS of Sc adatom on (c) pristine graphene and (d) B-codoped graphene. Blue solid lines: majority spin; red dot lines: minority spin.}
\label{figure2}
\end{figure}

We now investigate the magnetic property of the codoped graphene system. In Fig.~\ref{figure1}(b), one can find that the resulting magnetic moments of Sc, Ti, V,and Cr adatoms on graphene decreases when B atoms are codoped, while those of Mn, Fe, Co, and Ni adatoms increase after B codoping. In below, we take Fe and Sc atoms as examples to analyze the variance of the magnetic moments after B codoping. Figure~\ref{figure2} displays the orbital-resolved density of states of Fe and Sc adatoms on pristine and B doped graphene, respectively. Compared with the pristine graphene, the \textit{p}-doped B-substituted graphene weakens the hybridization effect of lowering the 3d orbital energies, therefore hindering charge transfer from 4s electrons to 3d orbitals. This makes the magnetic moments of the transition metal adatoms smaller (i.e., Sc, Ti, V) or larger (i.e., Fe, Co, Ni) than those on pristine graphene (see Fig. \ref{figure1}), while Cr and Mn atoms contribute their 4s electrons to the \textit{p}-doped graphene, making its magnetic moment exactly equal to 5 and 6 $\rm{\mu_B}$, respectively.
The orbital-resolved density of states presented in Fig.~\ref{figure2} further confirm this physical mechanism: When Fe atom is adsorbed on pristine graphene (see Fig.~\ref{figure2}(a)), electrons fully occupy five spin-up 3d orbitals and three spin-down 3d orbitals; After B codoping, electrons fully occupy five spin-up 3d orbitals but only two spin-down 3d orbitals (see Fig.~\ref{figure2}(b)), resulting in the increase of the magnetic moment from 2 $\rm{\mu_B}$ to 3 $\rm{\mu_B}$. For Sc adatoms, there are one spin-up 3d orbital fully occupied and two spin-up 3d orbitals partially occupied in the case of pristine graphene (see Fig.~\ref{figure2}(c)), while only one spin-up 3d orbital fully occupied and one spin-up 3d orbital partially occupied in the case of B-substituted graphene (see Fig.~\ref{figure2}(d)), leading to a smaller magnetic moment.

\begin{figure}
\includegraphics
[width=8.5cm,angle=0]{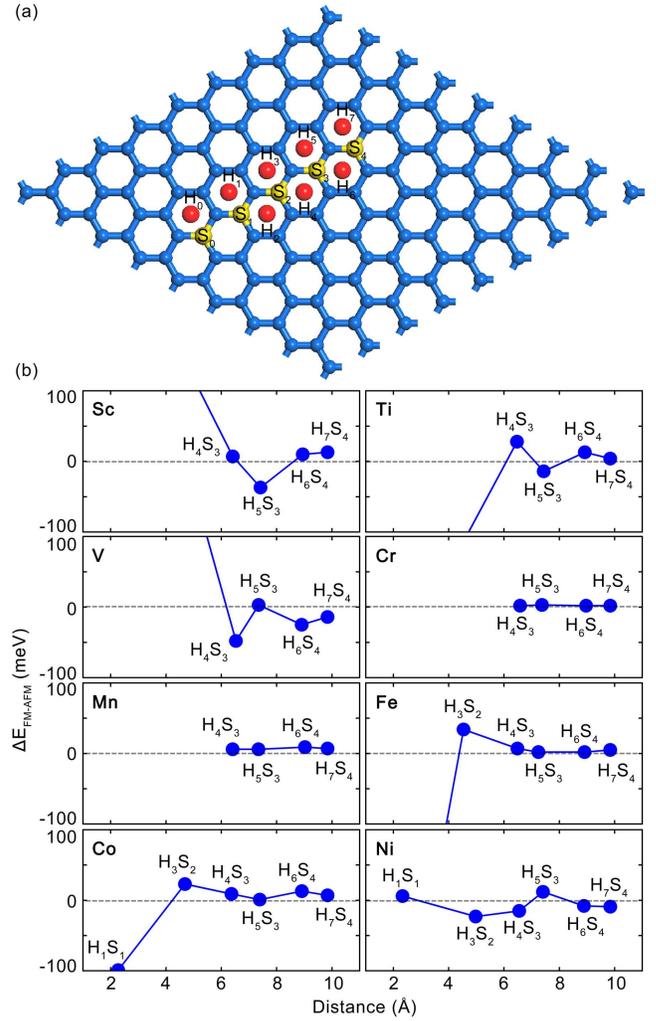}
\caption{ (color online). (a) A 8$\times$8 supercell of graphene. H$_0$$\sim$H$_7$ and S$_0$$\sim$S$_4$ represent sites for the transition metal adatoms and substitutional B dopants, respectively. (b) Magnetic coupling between two codopants pairs at different distances. One pair is fixed at H$_0$S$_0$ and the other moves from H$_1$S$_1$ to H$_7$S$_4$ as illustrated in (a).
}
\label{figure3}
\end{figure}

The above results show that the codoping approach can indeed stabilize the dilute distribution of metal adatoms with B-substitution in graphene, and the magnetic moments of Mn, Fe, Co and Ni significantly increase while those of Sc, Ti, V and Cr remain relatively large. Now we further study the magnetic interactions between two codopants (metal atom and B) pairs by setting them in a 8$\times$8 graphene supercell. As displayed in Fig.~\ref{figure3}(a), one metal adatom is located at $H_0$ site pairing with a B dopant at $S_0$ site, and the second metal adatom moves from $H_1$ to $H_7$ accompanying with the movement of its nearest-neighbor B codopant. Figure~\ref{figure3}(b) summarizes the energy difference between ferromagnetic and antiferromagnetic states of the two transition metal atoms at given separations. We find that the $1^{\rm st}$-$3^{\rm rd}$ nearest-neighbour configurations are unstable for Cr and Mn adatoms, because the attraction between two Cr or Mn adatoms is  surprisingly large to form clusters even though their corresponding adsorption energies increase after codoping with B atoms. The $2^{\rm nd}$/$3^{\rm rd}$ nearest-neighbour configurations of Sc, Ti, V and the $2^{\rm nd}$ nearest-neighbour configurations of Fe, Co, Ni are also unstable, with adatoms always tending to form the $1^{\rm st}$ nearest-neighbour configuration.

The strong ferromagnetism or anti-ferromagnetism coupling at the $1^{\rm st}$ nearest-neighbour distance for Sc, Ti, V, Fe and Co is closely related to the direct exchange interaction because their $1^{\rm st}$ nearest-neighbour distances are approximately equal to the $1^{\rm st}$ nearest-neighbour distances in their respective bulk crystals. While Ni adatoms exhibit weaker coupling energy because of their small magnetic moments. However, most of them present Ruderman-Kittel-Kasuya-Yoshida (RKKY)-like long-range magnetic interaction, except Cr and Mn that display paramagnetism. One can observe that only V-B and Ni-B codoped graphene systems exhibit ferromagnetism at some longer adatom-adatom distances. This indicates the potential long-range ferromagnetic order in these systems.

\begin{figure}
\includegraphics
[width=8.5cm,angle=0]{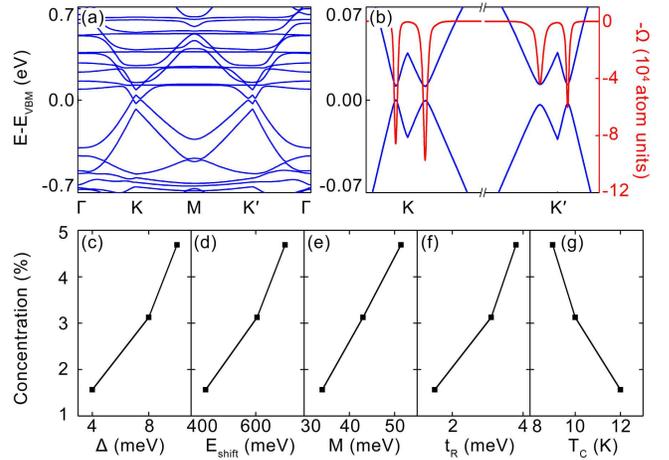}
\caption{ (color online).  (a) Full band structure of 6 Ni-B codoping pairs in the 8$\times$8 supercell of graphene along high symmetry lines. (b) Zoom in the bands around K and K$^{\prime}$ Dirac points. Red line represents the Berry curvature. (c) The band gap $\Delta$, (d) shift of valence band maximum from Fermi level $E_{\rm{shift}}$, (e) exchange field $M$, (f) Rashba spin-orbit coupling strength $t_R$ and (g) simulated Curie temperature $\rm{T_C}$ as a function of Ni-B pairs concentration.
}
\label{figure4}
\end{figure}

\textit{Ni-B codoping---.} From above analysis, we find that both V-B and Ni-B codoping may realize diluted ferromagnetism in graphene. Then we study whether the QAHE can be realized in V-B and Ni-B codoped graphene. Our further calculations show that the introduction of spin-orbit coupling in V-B codoped graphene does not open a bulk gap in any codoping concentration. Therefore, hereinbelow we only consider the Ni-B codoped graphene. We first investigate the band structures of Ni-B codoped graphene at different codoping concentrations. By codoping two, four and six Ni-B pairs in a 8$\times$8 supercell of graphene, one can get corresponding concentrations of 1.6\%, 3.1\% and 4.7\% respectively.
Due to the equal probability of different sublattices in graphene to be substituted experimentally, the amount of B dopants in A/B-sublattitce is set to be identical.
Figure~\ref{figure4}(a) displays band structure of codoping six Ni-B pairs in the 8$\times$8 supercell of graphene, with bulk band gaps opening near valleys K and K$^{\prime}$ when spin-orbit coupling is invoked. For three different codoping concentrations mentioned above, the resulting bulk gaps are respectively 4, 8, and 10 meV [see Fig.~\ref{figure4}(c)], suggesting the gap tunability by controlling the codoping concentration. Another obvious finding is that Fermi-levels lie outside gaps for all concentrations and the shift of valance band maximum from Fermi level enlarges with increase of doping concentration as plotted in Fig.~\ref{figure4}(d) because of \textit{p}-doping effect of B atoms. To realize the insulating nature, artificial adjustment is required to tune Fermi-levels into bulk gaps, e.g. by applying a gate voltage or tuning the ratio between B and Ni dopants.

So far, we show that Ni-B codoped graphene can form long-range ferromagnetism and the presence of spin-orbit coupling can further open up a band gap. In below, we investigate whether such a band gap can harbour the QAHE via Chern number calculation, which can be obtained by integrating Berry curvatures of the occupied valence bands using the expression~\cite{Berry1,Berry2}
\begin{eqnarray}
  \Omega(\bm{k})=-\sum_{n}f_n\sum_{n'\neq n} \dfrac{2 {\rm{Im}} \langle{\psi_{n{k}}|v_x|\psi_{n'{k}}\rangle} \langle{\psi_{n'{k}}|v_y|\psi_{n{k}}\rangle}}{(E_{n'}-E_n)^2}, \nonumber
\end{eqnarray}
where $n$, $E_n$ and $\psi_{nk}$ are the band index, eigenvalue and eigenstate of the $n$-th band , respectively. $v_{x,y}$=$\partial E/\partial k_{x,y}$ are velocity operators along $x$ and $y$ directions within the film plane, and $f_n$=1 for all $n$ bands below the band gap. Our calculation finds that this gap can host the QAHE. As an example, Fig.~\ref{figure4}(b) displays the Berry curvature distribution along high symmetry lines with large negative peaks appearing near valleys K and K$^\prime$ and vanishing elsewhere, demonstrating a nonzero Hall conductance. The Chern number can be obtained by integrating the Berry curvatures $\Omega(\bm{k})$ over the first Brillouin zone using the equation
\begin{eqnarray}
  \mathcal{C}=\dfrac{1}{2\pi}\int_\text{BZ}d^2k\Omega(\bm{k}), \nonumber
\end{eqnarray}
\Blue{which is numerically calculated to be $\mathcal{C}=2$ in our study.} This guarantees the formation of QAHE in Ni-B codoped graphene.

From a former theory~\cite{Gra-QAHE1}, it is known that both ferromagnetic exchange field ($M$) and Rashba spin-orbit coupling ($t_R$) play crucial roles in realizing graphene-based QAHE. By fitting our band structures around valleys K and K$'$, we can extract the corresponding parameters $M$ and $t_R$. Figures~\ref{figure4}(e) and \ref{figure4}(f) display the dependence of exchange field and Rashba spin-orbit coupling on the codoping concentration. One can see that with increasing doping concentration, both $M$ and $t_R$ raise almost linearly, which agrees with the band gap increasing from our first-principles calculations.

\textit{Curie temperature---.} The experimental observation temperature is determined by the lower limit of bulk band gap and ferromagnetic Curie temperature $\rm{T_C}$. We calculate Curie temperature ($\rm{T_C}$) by using Monte Carlo method~\cite{MC1,MC2} as applied to the diluted magnetic semiconductors within the classical Heisenberg model:
\begin{eqnarray}\label{Heisenburg}
H=&-& \sum_{i,j}J_{ij} {\mathbf{S}}_{i}{\cdot}{\mathbf{S}}_{j} \nonumber
\end{eqnarray}
where $J_{ij}$ is the magnetic coupling constant between moments $i$ and $j$, and $\mathbf{S}_{i}$ is a unit vector representing the direction of spin $i$. The magnetic coupling strength $J$ for the pair of magnetic atoms is obtained from energy difference between antiferromagnetic and ferromagnetic configurations ($J=E_{\rm{AFM}}-E_{\rm{FM}}$). The thermodynamic magnetization per atom and the susceptibility are respectively calculated by
\begin{eqnarray}\label{MagAndChi}
&m(T)=\langle[(\sum_i\mathbf{S}_i^x)^2+(\sum_i\mathbf{S}_i^y)^2+(\sum_i\mathbf{S}_i^z)^2]^{1/2}\rangle/N \nonumber \\
&\chi=N(\langle{m^2}\rangle-\langle{m}\rangle^2)/T, \nonumber
\end{eqnarray}
where $N$ is the number of magnetic atoms. In Monte Carlo simulation, we consider a $12\times12$ graphene supercell and average the normalized magnetization and susceptibility by collecting 10,000 configurations with randomly distributed Ni-B pairs. At a given temperature, the first 10,000 Monter Carlo steps are used for relaxation, and thermodynamic quantities are calculated in the following 10,000 steps. By analyzing the simulated susceptibility $\chi$ as a function of temperature for graphene with 1.6\%, 3.1\% and 4.7\% Ni-B codoping, $\rm{T_C}$ is estimated to be 12K, 10K and 9K respectively, which are highlighted by susceptibility peaks. Figure \ref{figure4}(g) displays that the ferromagnetic Curie temperature $\rm{T_C}$ doesn't grow with the codoping concentration because of the dominating antiferromagnetic state at the $\rm{H_1S_1}$ configuration as displayed in Fig. \ref{figure3}(b).

\textit{Conclusions---.} In summary, we systematically investigate the adsorption of 3d transition metal atoms from Sc to Ni on B-doped graphene using first-principles calculation methods. Comparing with the adsorption on pristine graphene, we find that the 3d transition metal atoms have larger adsorption energy on B-codoped graphene, thus significantly suppressing adatom migration and clustering. After investigating the magnetic states, we find that most exhibit RKKY magnetic fluctuations, but only V and Ni can form long-range ferromagnetism on B-doped graphene. Moreover, Berry curvature calculation confirms that only Ni-B codoped graphene can open a band gap to harbour the QAHE, and the nontrivial band gap can be tuned by the adsorption concentration. The estimated ferromagnetic Curie temperature can reach over 10 Kelvin for various codoping concentrations.

\textit{Acknowledgements.---} This work was financially supported by NNSFC (Grant No. 11474265), the China Government Youth 1000-Plan Talent Program, Fundamental Research Funds for the Central Universities (WK3510000001 and WK2030020027), and the National Key R \& D Program (Grant No. 2016YFA0301700). The Supercomputing Center of USTC is gratefully acknowledged for the high-performance computing assistance.

\end{document}